\documentclass[prc,amsmath,twocolumn,showkeys,showpacs,superscriptaddress]{revtex4}

\usepackage{gensymb}
\usepackage{graphicx,color}
\usepackage{amssymb}
\usepackage{enumerate}
\usepackage{verbatim}
\usepackage{natbib}

\begin{document}
  \newcommand {\nc} {\newcommand}
  \nc {\Sec} [1] {Sec.~\ref{#1}}
  \nc {\IR} [1] {\textcolor{red}{#1}} 

\title{Explicit inclusion of nonlocality in $(d,p)$ transfer reactions}

\author{L.~J.~Titus}
\affiliation{National Superconducting Cyclotron Laboratory, Michigan State University, East Lansing, MI 48824, USA}
\affiliation{Department of Physics and Astronomy, Michigan State University, East Lansing, MI 48824-1321}
\author{F.~M.~Nunes}
\affiliation{National Superconducting Cyclotron Laboratory, Michigan State University, East Lansing, MI 48824, USA}
\affiliation{Department of Physics and Astronomy, Michigan State University, East Lansing, MI 48824-1321}
\author{G.~Potel}
\affiliation{National Superconducting Cyclotron Laboratory, Michigan State University, East Lansing, MI 48824, USA}
\affiliation{Lawrence Livermore National Laboratory L-414, Livermore, CA 94551, USA}

\date{\today}


\begin{abstract}
\begin{description}

\item[Background:] Traditionally, nucleon-nucleus optical potentials are made local for convenience.  In recent work we studied the effects of including nonlocal interactions explicitly in the final state for (d,p) reactions, within the distorted wave Born approximation. 
\item[Purpose:] Our goal in this work is to develop an improved formalism for nonlocal interactions that includes deuteron breakup and to use it to study the effects of including nonlocal interactions in transfer (d,p) reactions, in both the deuteron and the proton channel.  
\item[Method:] We extend the finite-range adiabatic distorted wave approximation to include nonlocal nucleon optical potentials. We apply our method to 
$(d,p)$  reactions on $^{16}$O, $^{40}$Ca, $^{48}$Ca, $^{126}$Sn, $^{132}$Sn, and $^{208}$Pb at $10$, $20$ and $50$ MeV. 
\item[Results:] We find that nonlocality in the deuteron scattering state reduces the amplitude of the wave function in the nuclear interior, and shifts the wave function outward. In many cases, this has the effect of increasing the transfer cross section at the first peak of the angular distributions. This increase was most significant for heavy targets and for reactions at high energies. 
\item[Conclusions:] Our systematic study shows that, if only local optical potentials are used in the analysis of experimental $(d,p)$ transfer cross sections, the extracted spectroscopic factors may be incorrect by up to $40\%$ due to the local approximation. 
\end{description}
\end{abstract}



\maketitle
\section{Introduction}

Nuclear reactions play a key role in the study of nuclei away from stability. Experimentally, single-nucleon transfer reactions involving deuterons provide an exceptional tool to study the single-particle shell structure of nuclei \cite{Schmitt_prl2012, Jones_prc2011}. These types of reactions are also used in extracting important astrophysical quantities and information relevant to stockpile stewardship \cite{Kozub_prl2012, Langer_prl2014, Cizewski2012}. Theoretically, these reactions are attractive as they can be cast into a three-body problem composed of $n+p+A$.   

In recent years, there have been many efforts to improve the description of deuteron induced transfer reactions. Studies benchmarking some of the simple and exact methods available have shown serious limitations in even the most advanced approaches \cite{Nunes_prc2011, Upadhyay_prc2012, Deltuva_prc2013}. Although a large body of work has been done to solve the three-body scattering problem accurately, other sources of uncertainty are still limiting the reliability of predicted cross sections \cite{Lovell_jpg2015}. The focus of this work will be on the uncertainties associated with the ambiguities in the optical potentials, particularly the effect of nonlocality on transfer reaction observables.

For many decades it has been known that the optical model potential for nucleon-nucleus scattering is nonlocal \cite{Bell_prl1959}. The sources of nonlocality in these effective potentials arise predominantly from anti-symmetrization \cite{Canton_prl2005, Rawitscher_prc1994,Fraser_epja2008} and channel couplings \cite{Fraser_epja2008, Rawitscher_npa1987,Feshbach_ap1958, Feshbach_ap1962}. Over the years, a numbers of semi-microscopic approaches were developed to construct optical potentials and have been applied to a variety of problems in nuclear reactions (e.g.  \cite{Bauge_prc1998, Bauge_prc2001}). One recent example is the dispersive optical model (DOM) introduced by Mahaux and Sartor \cite{Mahaux_anp1991} which provides a  link between nuclear reactions and nuclear structure through a dispersion relation. This approach has been recently revisited by the St. Louis group, with a first realization of the nonlocal DOM potential for nucleon-$^{40}$Ca \cite{Mahzoon_prl2014}. 

There are only a few global nonlocal optical potentials available \cite{Perey_np1962, Tian_ijmpe2015, Giannini_ap1976, Giannini_ap1980}. All of these nonlocal potentials are energy independent, with the parameters constrained by elastic scattering data, and in some cases polarization data. While all of these potentials are energy-independent, on physical grounds the nonlocal potential should have an energy dependence due to the effects of channel couplings. These potentials are all of the Frahn-Lemmer form \cite{Frahn_nc1957}, which consists of a Woods-Saxon form factor multiplied by a Gaussian nonlocality term. In this work we will take the potential of Perey-Buck \cite{Perey_np1962}.

While both the knowledge of nonlocality in the nuclear potential and nonlocal optical potentials fitted to elastic scattering have been around for over half a century, they have never received widespread use. Instead, the preferred method was to use strongly energy-dependent local potentials, where the energy dependence was assumed to incorporate the effects of nonlocality. Throughout the years, numerous global parameterizations of local phenomenological optical potentials have been developed (e.g. \cite{koning_np2003, ch89, bg69}).  

Ever since the first studies of nonlocality, it has been known that, depending on the observable, replacing a nonlocal potential with an energy dependent local potential may be  insufficient, even if the two potentials reproduce the same elastic scattering. This is because a nonlocal potential causes a reduction of the wave function within the nuclear interior, a phenomenon known as the Perey effect \cite{Austern_pr1965, Fiedeldey_np1966}. For many decades after, nonlocality was taken into account approximately by multiplying the local wave functions by an approximate correction factor \cite{Austern_pr1965, Fiedeldey_np1966}. This approach was recently tested for $(p,d)$ transfer reactions \cite{Titus_prc2014} where nonlocality was considered in the entrance channel only.  As compared to using local potentials, the explicit inclusion of nonlocality in the entrance channel increased the transfer cross section by up to $35$\%. In all cases the correction factor moved the local angular distribution in the direction of the nonlocal angular distribution, but could  not reproduce the full effects. That study suggested that  nonlocality needs to be explicitly included in the calculations  for obtaining reliable quantitative results. A subsequent study, using the DOM potential  \cite{Mahzoon_prl2014}, emphasized this point \cite{Ross_prc2015}. Therein, differences in the first peak of the transfer angular distribution of up to $50$\% were seen. 

The above mentioned studies, investigating the effect of nonlocality on $(p,d)$ reactions \cite{Titus_prc2014,Ross_prc2015}, used the Distorted Wave Born Approximation (DWBA) and focused on including nonlocality only in the proton channel. This work upgrades the original studies in two ways: i) it replaces the DWBA description of the reaction by including deuteron breakup to all orders within the three-body finite-range Adiabatic Distorted Wave Approximation (ADWA) \cite{Johnson_npa1974} and ii) it considers nonlocality in all effective nucleon-target interactions, both in the entrance and exit channel. It thus provides a complete picture of the role of nonlocality.

Although DWBA has been a very popular method to analyze transfer reactions, it has long been understood that deuteron induced reactions are better described when deuteron breakup is included explicitly in the formalism. One then needs to think in terms of a three-body $n+p+A$ system, dependent on nucleon-target interactions, which can be solved exactly through Faddeev approaches (e.g. \cite{Deltuva_prc2013}).  Benchmark studies have shown that other approaches such as the Continuum Discretized Coupled Channel (CDCC) method or the adiabatic approach (ADWA) \cite{Nunes_prc2011, Upadhyay_prc2012} provide a reliable description of the transfer cross sections for the energies of interest. For its computational feasibility we choose to use ADWA for the framework of the current study.  ADWA was originally developed and implemented for local interactions only. In this work we extend the method to include nonlocal nucleon-target interactions.

While nowadays differential equations with nonlocal interactions are easy to solve computationally, it is still attractive to transform the nonlocal problem into a local one. Recently, a method was proposed to effectively incorporate nonlocality in the deuteron channel through an energy shift \cite{Timofeyuk_prl2013, Timofeyuk_prc2013}. Traditionally, within the adiabatic distorted wave approximation (ADWA) \cite{Johnson_npa1974}, one uses nucleon optical potentials evaluated at half the deuteron energy to construct the effective adiabatic deuteron potential. In \cite{Timofeyuk_prl2013, Timofeyuk_prc2013} it is suggested that nonlocality in the deuteron channel may be effectively included by  shifting the energy at which the local nucleon potentials are evaluated. In this study, we also compare our results with those obtained in \cite{Timofeyuk_prc2013}.

In this work, we investigate the inclusion of full nonlocality in $(d,p)$ transfer reactions. Nonlocality is included explicitly and consistently, within the ADWA. We reformulate the ADWA in terms of nonlocal nucleon optical potentials and compare the results obtained using nonlocal interactions with the corresponding local interactions. Our systematic study includes $(d,p)$ reactions on $^{16}$O, $^{40}$Ca, $^{48}$Ca, $^{126}$Sn, $^{132}$Sn, and $^{208}$Pb at deuteron energies of $10$, $20$, and $50$ MeV. 
The paper is organized in the following way. 
In Sec. \ref{theory} we briefly describe the necessary theory, followed by the numerical details in Sec. \ref{numerical}. The results are presented in Sec. \ref{results}. We compare our results with those from other methods and explore specific sensitivities in Sec. \ref{Comparisons}. Finally, in Sec. \ref{conclusions}, conclusions are drawn.


\section{Theoretical framework}
\label{theory}

We are interested in describing the reaction A(d,p)B where the final nucleus $B$ is in a bound state.
Given the importance of deuteron breakup in deuteron induced reactions, we
begin with  a three-body Hamiltonian for the $n+p+A$ system, 

\begin{eqnarray}
\mathcal{H}_{3B}=T_R+T_r+U_{nA}+U_{pA}+V_{np}\;.
\label{eq-h3b}
\end{eqnarray}

\noindent Here $T_R$ and $T_r$ are the kinetic energy operators for the center of mass motion and $n-p$ relative motion, respectively,  $V_{np}$ is the neutron-proton interaction, while $U_{pA}$  and $U_{nA}$ are the proton-target and neutron-target interactions. In principle, these interactions can be either local or nonlocal. The wave function $\Psi(\textbf{r},\textbf{R})$ describes a deuteron incident on a nucleus $A$ and is a solution to the equation $\mathcal{H}_{3B}\Psi=E\Psi$, where $E=E_d-\epsilon_d$ is the energy of the system, $E_d$ is the incident deuteron kinetic energy in the center of mass frame, and $\epsilon_d$ is the deuteron binding energy. The vectors locating the proton and neutron, $\textbf{R}_p$ and $\textbf{R}_n$, are taken relative to the center of mass of the target $A$, and the vectors $\textbf{r}$ and $\textbf{R}$ are given by $\textbf{R}=\left(\textbf{R}_p+\textbf{R}_n \right)/2$ and $\textbf{r}=\textbf{R}_p-\textbf{R}_n$.

The exact T-matrix for the $A(d,p)B$ reaction can be written in the post form as:

\begin{eqnarray}\label{eq-tmatrix}
T=\langle \phi_{nA}\chi^{(-)}_{pB}|V_{np}+\Delta|\Psi^{(+)}\rangle,
\label{eq-tmatrix}
\end{eqnarray}

\noindent where $\phi_{nA}$ is the neutron-target bound state, and $\chi^{(-)}_{pB}$ is the proton scattering state in the exit channel distorted by $U^*_{pB}$. The remnant term $\Delta$ is negligible for all but light targets.

When computing transfer as in Eq.(\ref{eq-tmatrix}), it is seen that $\Psi^{(+)}$ is only needed within the range of the $V_{np}$ interaction. In these circumstances, a basis such as the Weinberg states \cite{Weinberg_pr1963}, which forms a complete and orthonormal set only within the range of the $V_{np}$ interaction, is sufficient for expanding the three-body wave function. This was the realization by Johnson and Tandy \cite{Johnson_npa1974} and gave rise to the finite range adiabatic model (ADWA). First, one expands $\Psi^{(+)}$ in Weinberg states, 

\begin{equation}\label{eq-psi}
\Psi^{(+)}(\textbf{r},\textbf{R})=\sum_{i=0}^{\infty}\Phi_{i}(\textbf{r})X_{i}(\textbf{R}).
\end{equation}

\noindent Each Weinberg state is defined by 

\begin{eqnarray}
\left(T_{\textbf{r}}+\alpha_iV_{np} \right)\Phi_i(\textbf{r})=-\epsilon_d\Phi_i(\textbf{r}),
\end{eqnarray}

\noindent where $\epsilon_d$ is the deuteron binding energy, $\alpha_i$ are the eigenvalues, and each state is normalized by $\langle \Phi_i|V_{np}|\Phi_j\rangle = -\delta_{ij}$. 

Introducing the expansion Eq.(\ref{eq-psi}) for $\Psi^{(+)}$ into the three-body Schr\"odinger equation produces a coupled channel equation for the $X_i(\textbf{R})$ \cite{Laid_prc1993}. It has been shown \cite{Pang_prc2013} that, to a very good approximation, the expansion in Eq.(\ref{eq-psi}) can be truncated after the first term, generating an optical model like equation with an effective adiabatic potential for the deuteron that contains the effect of breakup to all orders. This adiabatic potential is defined by \cite{Johnson_npa1974}:

\begin{eqnarray}
U_{AD}=-\langle \Phi_0(\textbf{r})|V_{np}\left(U_{nA}+U_{pA} \right)|\Phi_{0}(\textbf{r})\rangle,
\label{eq-pot}
\end{eqnarray}
with $\phi_0$ being the first Weinberg eigenstate.

In the way Eq.(\ref{eq-pot}) is written, no assumption is made on the locality or nonlocality of the nucleon-target interactions.
When local potentials are used, the arguments of $U_{pA}$ and $U_{nA}$ are the distances from each of the nucleons to the target, $\textbf{R}_{p}$ and $\textbf{R}_{n}$ respectively. After integrating over $\textbf{r}$, $U_{AD}$ becomes a function of $\textbf{R}$. In that case the three-body equation for the incoming deuteron scattering state becomes:

\begin{eqnarray}
\left[T_{\textbf{R}}+U^{Loc}_{AD}(\textbf{R}) + U_{so}(R) + U_{Coul}(R)-E_d \right]X_{ad}(\textbf{R})=0, \nonumber \\
\label{eq-adwa}
\end{eqnarray}

\noindent This is the method that has been implemeneted and used for several studies (e.g. \cite{Nunes_prc2011}).

We now want to extend the method to nonlocal interactions. For the sake of simplicity, we will not include nonlocality in the spin-orbit term nor in the Coulomb interaction, and focus on the effect on the nuclear central interaction which is the dominant force.
The nonlocal operators $U_{pA}$ and $U_{nA}$ acting on $\Psi^{(+)}(\textbf{r},\textbf{R})$ are given by

\begin{eqnarray}
&\hat{U}_{pA}&\Psi^{(+)}(\textbf{r},\textbf{R})= 8 \times \nonumber \\
&&\int d\textbf{s} \; U_{pA}(\textbf{R}_{p},\textbf{R}_{p}+2\textbf{s})\Psi^{(+)}(\textbf{r}+2\textbf{s},\textbf{R}+\textbf{s}) \nonumber \\ 
&\hat{U}_{nA}&\Psi^{(+)}(\textbf{r},\textbf{R}) = 8 \times \\
&&\int d\textbf{s} \; U_{nA}(\textbf{R}_{n},\textbf{R}_{n}+2\textbf{s})\Psi^{(+)}(\textbf{r}-2\textbf{s},\textbf{R}+\textbf{s}), \nonumber
\end{eqnarray}

\noindent where $\textbf{s}=\textbf{R}'-\textbf{R}$. The ADWA method for nonlocal optical potentials is somewhat similar to the original ADWA method. The original  Eq.(\ref{eq-adwa}) becomes:
\begin{eqnarray}
\left[T_{\textbf{R}}+U_{so}(R) + U_{Coul}(R)- E_d \right]X^{NL}_{ad}(\textbf{R})=S^{NL}_{AD}(\textbf{R}), \nonumber \\
\label{eq-nladwa}
\end{eqnarray}
and the main difference is now that instead of a local adiabatic potential, we need to deal with a source term that is given by:

\begin{eqnarray}
&\phantom{=}&S_{AD}^{NL}(\textbf{R})=8\int d\textbf{r}d\textbf{s}\Phi_o(\textbf{r})V_{np}(\textbf{r}) \nonumber \\
&\phantom{=}& \left[U_{pA}(\textbf{R}_{p},\textbf{R}_{p}+2\textbf{s})\Phi_0(\textbf{r}+2\textbf{s})X^{NL}_{ad}(\textbf{R}+\textbf{s}) \right. \\
&\phantom{=}& \ \left. + U_{nA}(\textbf{R}_{n},\textbf{R}_{n}+2\textbf{s})\Phi_0(\textbf{r}-2\textbf{s})X^{NL}_{ad}(\textbf{R}+\textbf{s}) \right]. \nonumber
\label{eq-source}
\end{eqnarray}
Starting from these equations we have applied the partial wave decomposition, performed the necessary angular momentum coupling and designed an effective method for the angular integrations. Details can be found in Appendix A.

\section{Numerical details}
\label{numerical}

We begin by defining the nonlocal interactions needed for solving this three-body problem Eq.(\ref{eq-h3b}). The Perey-Buck nonlocal potential was used for all proton and neutron optical potentials, both in the entrance and exit channel. For the final bound states, we take the real part of the Perey and Buck interaction, and adjust the depth to reproduce the experimental binding energy of the system.  The method used to solve the nonlocal equations is detailed in \cite{Titus_prc2014}. The NN interaction is a central Gaussian which reproduces the binding energy and radius of the deuteron ground state, as in \cite{Moro_prc2009}. 

To understand the role of nonlocality, it is important to compare our results with a local calculation that has identical constrains. We thus impose, for the scattering, that the local nucleon-target interactions used reproduce the elastic scattering obtained using Perey-Buck at the relevant energies. These local phase equivalent (LPE) potentials are found by fitting the elastic scattering distributions using the code \textsc{SFRESCO} \cite{fresco}. In the proton channel, the procedure is straightforward: $(p,p)$ elastic scattering is calculated at the relevant energy using the Perey-Buck nonlocal potential, and the LPE potential is found by fitting this distribution.  In the deuteron channel more care is needed. We calculate $(n,n)$ and $(p,p)$ elastic scattering at half the deuteron energy using the Perey-Buck potential. We find LPE potentials for these elastic scattering distributions, and calculate the local adiabatic potential with the neutron and proton LPE potentials. The fact that we impose that our local $U_{nA}$ and $U_{pA}$ are phase equivalent to Perey-Buck does not guarantee that the local and nonlocal adiabatic potentials are phase equivalent, and in general this will not be the case. In fact, the adiabatic method only provides a good description of the three-body scattering wave function in the region of $V_{np}$ and therefore is not useful for calculating elastic scattering phase shifts.  

For the bound states, we calculate the nonlocal equation using a real Woods-Saxon form and the nonlocality parameter set to $\beta=0.85$. We also use a local spin-orbit interaction with the depth fixed at $6$ MeV. For each term we use a radius parameter of $r=1.25$ fm. The diffuseness was set to $a=0.65$ fm. The depth of the nonlocal real Woods-Saxon form is then adjusted to reproduce the physical binding energy. The corresponding local equation is obtained by settting $\beta=0$ and adjusting the local real Woods-Saxon depth to reproduce the binding energy.  

The bound and scattering states calculated using either nonlocal or local potentials are then introduced into the $(d,p)$ $T$-matrix, Eq. (\ref{eq-tmatrix}). Angular distributions are calculated for the targets under consideration for deuteron energies of $10$, $20$, and $50$ MeV, spanning a wide range of experimental interest. The radial wave functions are calculated using a step size of $0.01$ fm and a matching radius of $30$ fm. 
The nonlocal adiabatic potential is obtained on a radial grid of step $0.05$ fm and a linear interpolation is  used in the differential equation to calculate the adiabatic deuteron wave function. Converged cross sections contain partial waves up to $J=30$. 

\section{Results}
\label{results}

The effect of nonlocality on the neutron bound states and proton scattering states were studied extensively in previous work \cite{Titus_prc2014,Ross_prc2015}. Here, the same qualitative results were found. Concerning bound states, nonlocality reduces the amplitude within the nuclear interior and, due to the normalization condition, this produces an increase in the asymptotic normalization coefficient (ANC) when compared to the local counterpart. As to the proton scattering states, nonlocality also reduces the amplitude of the wave function within the nuclear interior as compared to the local counterpart, but has no effect on the asymptotic properties of the scattering wave functions by construction (phase equivalent).   What is new in this study is the effect of nonlocality in the deuteron scattering state. We begin by looking into the source term in Eq.(\ref{eq-source}).


\begin{figure}[h]
\begin{center}
\includegraphics[width=0.5\textwidth]{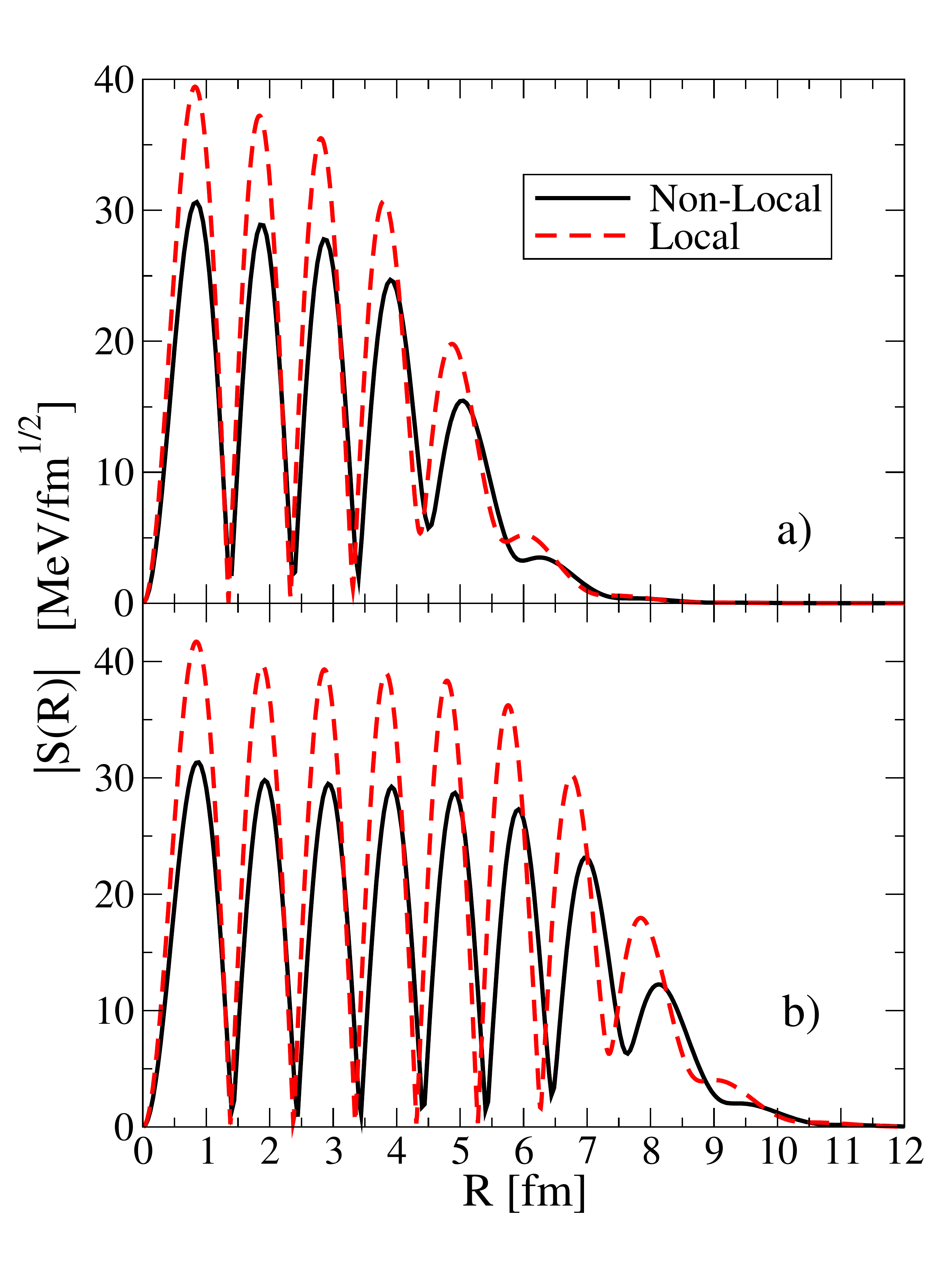}
\end{center}
\caption{Absolute value of the $d+A$ source term when nonlocal and local potentials are used. (a) $d+^{48}$Ca at $E_d=50$ MeV. (b) $d+^{208}$Pb at $E_d=50$ MeV. Both are for the $L=1$ and $J=0$ partial wave.}
\label{fig:Source_Comparison}
\end{figure}

\subsection{Source term in the deuteron scattering equation}
\label{DeuteronSource}

To get an idea of the effect of nonlocality on the adiabatic potential, in Fig. \ref{fig:Source_Comparison} we compare the source term $S^{NL}_{AD}(\textbf{R})$ with the local correponding quantity $U_{AD}^{Loc}(\textbf{R})X_{ad}(\textbf{R})$, as a function of the radial distance between the deuteron and the target. The top panel is for $d+^{48}$Ca and the bottom panel for $d+^{208}$Pb, both at a beam energy of $50$ MeV in the laboratory frame. The solid line corresponds to the nonlocal source term, while the dashed line is its local equivalent. The magnitude of the nonlocal source term is reduced compared to the local source term. This is related to the reduction of the scattering wave function in the nonlocal case, known as the Perey effect. It is also seen that the source term in the nonlocal case gets shifted outward relative to the local case. Both these effects imprint themselves on the adiabatic deuteron wave function discussed next.

\subsection{Deuteron Scattering Wave Functions}
\label{DeuteronWF}

\begin{figure}[h]
\begin{center}
\includegraphics[width=0.5\textwidth]{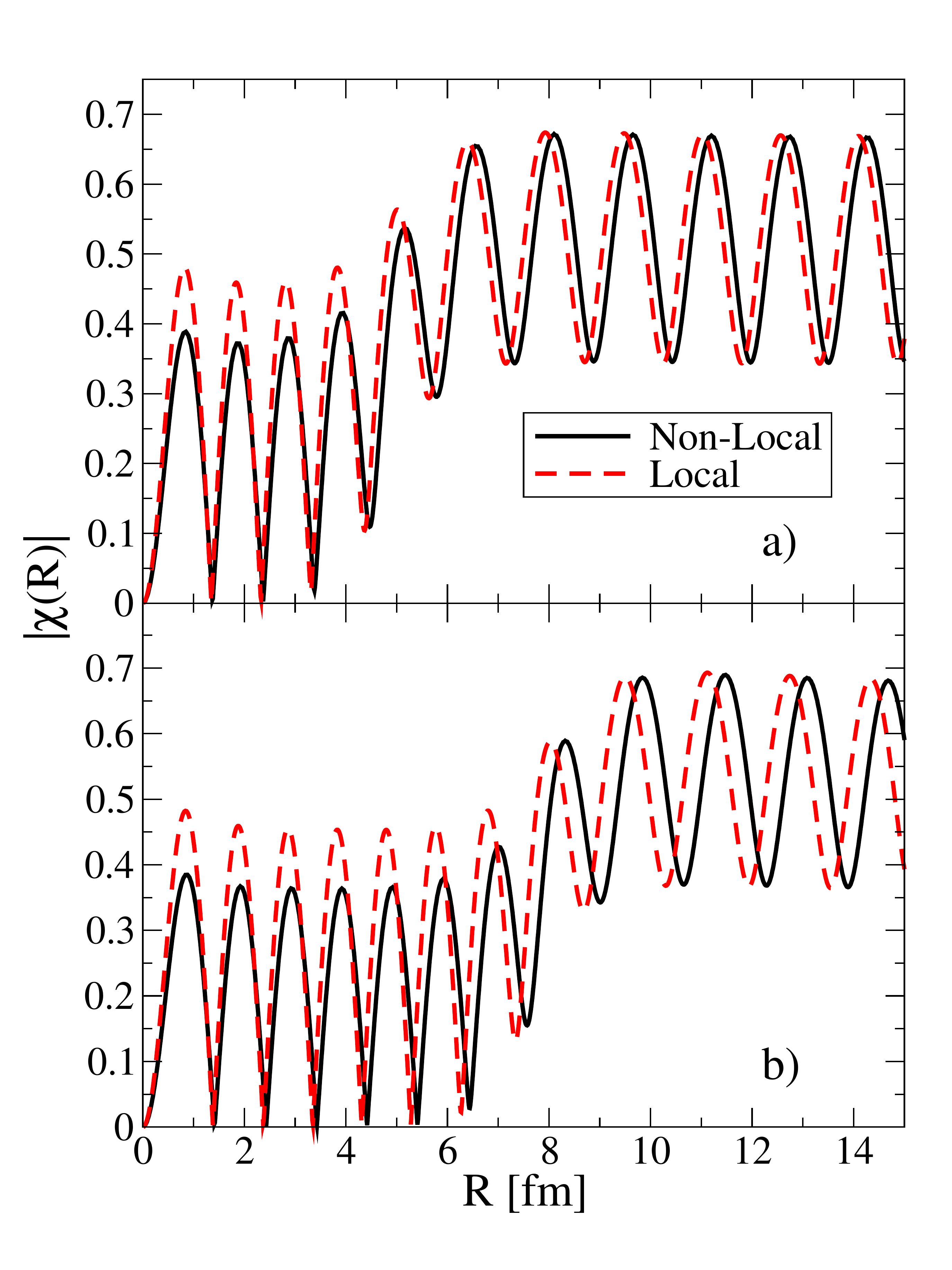}
\end{center}
\caption{Absolute value of the $d+A$ scattering wave function when nonlocal and local potentials are used. (a) $d+^{48}$Ca at $E_d=50$ MeV. (b) $d+^{208}$Pb at $E_d=50$ MeV. Both are for the $L=1$ and $J=0$ partial wave.}
\label{fig:dWF_Comparison}
\end{figure}

In Fig. \ref{fig:dWF_Comparison} we show the absolute values of the $d+A$ wave functions as a function of radius for the channel with deuteron-target angular momentum $L=1$ coupled to a total spin $J=0$. The solid line is obtained with the nonlocal adiabatic potential Eq.(\ref{eq-nladwa}) and the dashed line is obtained with the local Eq.(\ref{eq-adwa}). We show the same reactions as in Fig. \ref{fig:Source_Comparison}. As in the other cases, nonlocality produces a reduction of the deuteron wave function in the nuclear interior, but also a shift of the wave function towards the outside of the nucleus, since the nonlocal adiabatic potential is not phase equivalent to the local adiabatic potential.

\subsection{Transfer Cross Sections}
\label{Transfer}

\begin{figure}[h!]
\begin{center}
\includegraphics[width=0.5\textwidth]{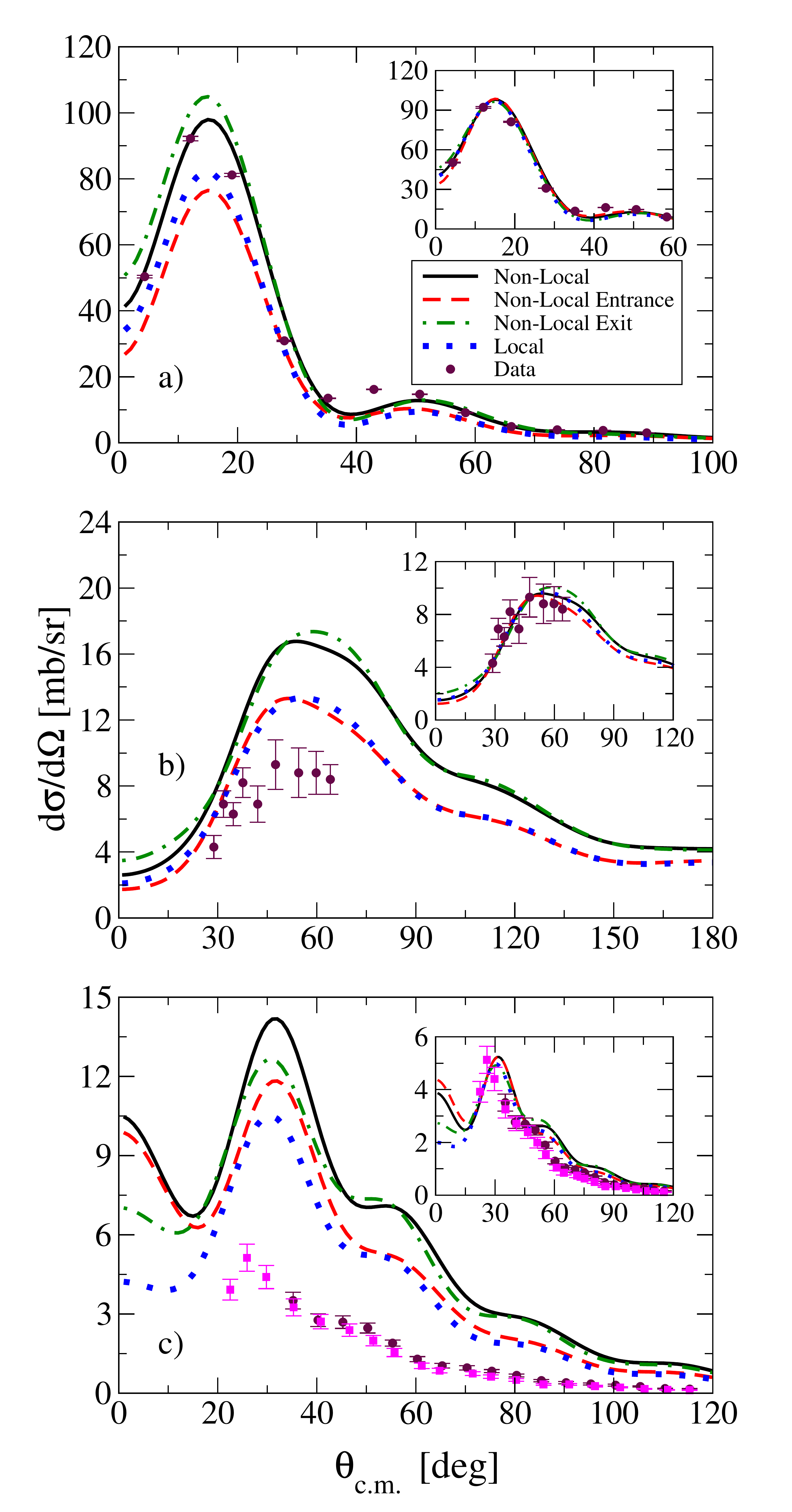}
\end{center}
\caption{Angular distributions for $(d,p)$ transfer cross sections as a function of scattering angle. The insets are the theoretical distributions normalized to the peak of the data distribution. (a) $^{48}$Ca$(d,p)^{49}$Ca at $E_d=10$ MeV with data \cite{Brown_npa1970} at $10$ MeV in arbitrary units. (b) $^{132}$Sn$(d,p)^{133}$Sn at $E_d=10$ MeV with data \cite{Jones_prc2011} at $E_d=9.4$ MeV. (c) $^{208}$Pb$(d,p)^{209}$Pb at $20$ MeV with data \cite{Hirota_npa1998} (Circles) and \cite{Seichert_aip1981} (Squares) at $E_d=22$ MeV.}
\label{fig:TransferCS_LowE}
\end{figure}

\begin{figure}[h]
\begin{center}
\includegraphics[width=0.5\textwidth]{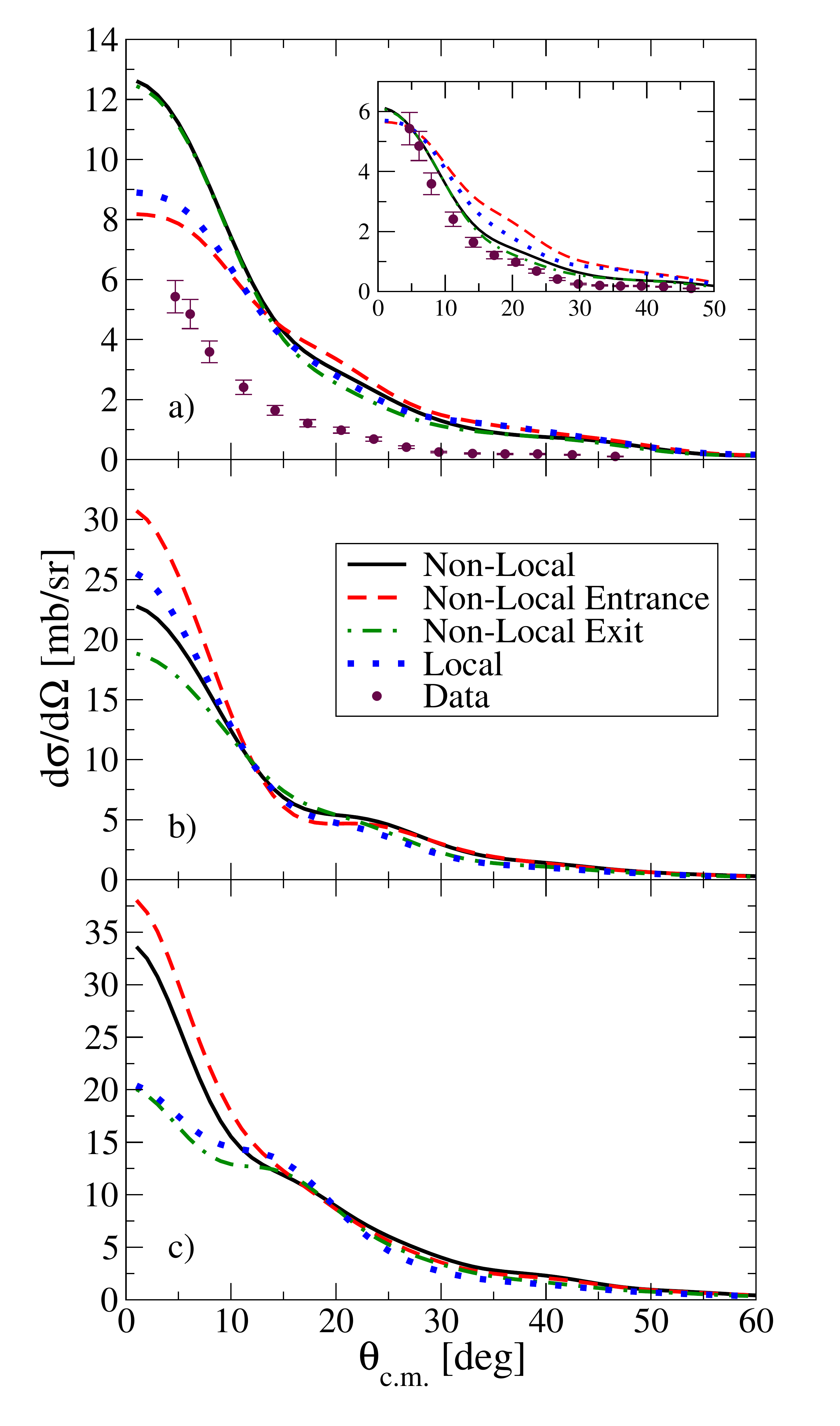}
\end{center}
\caption{Angular distributions for $(d,p)$ transfer cross sections as a function of scattering angle. The insets are the theoretical distributions normalized to the peak of the data distribution. (a) $^{48}$Ca$(d,p)^{49}$Ca at $E_d=50$ MeV with data \cite{Uozumi_npa1994} at $56$ MeV. (b) $^{132}$Sn$(d,p)^{133}$Sn at $E_d=50$ MeV. (c) $^{208}$Pb$(d,p)^{209}Pb$ at $50$ MeV.}
\label{fig:TransferCS_HighE}
\end{figure}

Next we turn to the (d,p) cross sections. In our analysis, we compute angular distributions for a wide variety of cases from $^{16}$O to $^{208}$Pb. Some examples to illustrate our results are shown in Fig. \ref{fig:TransferCS_LowE}  and Fig. \ref{fig:TransferCS_HighE}. An extensive list of all cases considered is shown in Tables \ref{Tab:Percent_Difference_10}, \ref{Tab:Percent_Difference_20} and \ref{Tab:Percent_Difference_50}. 

Fig. \ref{fig:TransferCS_LowE}  and Fig. \ref{fig:TransferCS_HighE} include the results of the full nonlocal calculation (solid black line), including nonlocality only in the deuteron channel (red dashed line), including nonlocality in the proton channel only (green dot-dashed line) and including local phase equivalent potentials as described in Sec. \ref{numerical} (blue dotted line). In Fig. \ref{fig:TransferCS_LowE} we present: (a) $(d,p)^{48}$Ca at $E_d=10$ MeV, (b)  $^{132}$Sn at $E_d=10$ MeV, and (c) $^{208}$Pb at $E=20$ MeV. The same cases are considered in Fig. \ref{fig:TransferCS_HighE}  but now at $E_d=50$ MeV.

Also shown are the data points when available. The data in Fig. \ref{fig:TransferCS_LowE}a was published in arbitrary units. Therefore, in this work, this set of data is normalized to the peak of the theoretical distribution generated by including nonlocality in both, the entrance and exit channels. All other data presented in this work has absolute units. 
It is important to note that the purpose of this work is not to describe the data. We do not expect the Perey and Buck optical potential developed in the sixties for $^{208}$Pb at intermediate energies to do well for a wide range of energies and targets. One should focus on the differences between the nonlocal and local transfer calculations under the constrain of the same physical input, namely that both the nonlocal and local nucleon optical potentials reproduce the exact same elastic scattering.

At low energy, the inclusion of nonlocality in the exit channel provides a significant enhancement of the cross section for all cases, resulting from the large effect of nonlocality in the neutron bound state. As mentioned before, the nonlocal wave functions have a larger  ANC and, at low energies, the reaction is only sensitive to this region. This results in a larger cross section for the nonlocal case (compare dotted with dot-dashed lines).

The nonlocality in the proton scattering state is only felt at the larger beam energies. At low energies the reaction is not sensitive to the short-range properties of the proton scattering wave function, so the fact that the wave function is reduced in the nuclear interior does not have a significant effect on the transfer cross section, as was seen in \cite{Titus_prc2014} and \cite{Ross_prc2015}. 

\begin{table}[h]
\centering
\begin{tabular}{|c|r|r|r|r|r|}
\hline
                &              &              & Nonlocal      & Nonlocal    &       \\
                &Final         & Nonlocal     & Entrance      & Exit        & Angle \\
$E_{lab}$        &Bound         & Relative     & Relative      & Relative    & of    \\
$10$ MeV        &State         & to Local     & to Local      & to Local    & Peak  \\
\hline
$^{16}$O$(d,p)$    & $1d_{5/2}$  & $27.2\%$    & $-3.0\%$     & $32.7\%$      & $26 ^\circ$  \\ 
$^{16}$O$(d,p)$    & $2s_{1/2}$  & $15.5\%$    & $0.2\%$      & $13.5\%$      & $0^\circ$    \\ 
$^{40}$Ca$(d,p)$   & $1f_{7/2}$  & $48.5\%$    & $11.4\%$     & $46.5\%$      & $39^\circ$   \\ 
$^{48}$Ca$(d,p)$   & $2p_{3/2}$  & $19.4\%$    & $-6.8\%$     & $27.8\%$      & $15^\circ$   \\ 
$^{126}$Sn$(d,p)$  & $1h_{11/2}$ & $36.9\%$    & $8.7\%$      & $26.9$        & $72^\circ$    \\  
$^{132}$Sn$(d,p)$  & $2f_{7/2}$  & $25.7\%$    & $-0.2\%$     & $30.1\%$      & $55^\circ$    \\ 
$^{208}$Pb$(d,p)$  & $2g_{9/2}$  & $52.5\%$    & $2.0\%$      & $47.3\%$      & $180^\circ$   \\ 
\hline
\end{tabular}
\caption{Percent difference of the $(d,p)$ transfer cross sections at the first peak when using nonlocal potentials in entrance and exit channels (1st column), nonlocal potentials in entrance channel only (2nd column), and nonlocal potentials in exit channel only (3rd column), relative to the local calculation with the LPE potentials, for a number of reactions occurring at 
$10$ MeV.}
\label{Tab:Percent_Difference_10}
\end{table}

\begin{table}[h]
\centering
\begin{tabular}{|c|r|r|r|r|r|}
\hline
                &      &                   & Nonlocal          & Nonlocal    &         \\
                &Final & Nonlocal          & Entrance          & Exit        & Angle   \\
$E_{lab}$        &Bound & Relative          & Relative          & Relative    & of   \\
$20$ MeV        &State & to Local          & to Local          & to Local    & Peak \\
\hline
$^{16}$O$(d,p)$   & $1d_{5/2}$  & $24.9\%$  & $2.6\%$ & $25.7\%$  & $0^\circ$ \\ 
$^{16}$O$(d,p)$   & $2s_{1/2}$  & $7.1\%$   & $-0.7\%$ & $6.0\%$   & $0^\circ$  \\ 
$^{40}$Ca$(d,p)$  & $1f_{7/2}$  & $43.3\%$  & $11.0\%$ & $34.1\%$          & $26^\circ$ \\ 
$^{48}$Ca$(d,p)$  & $2p_{3/2}$  & $14.9\%$  & $7.1\%$ & $12.2\%$           & $8^\circ$ \\ 
$^{126}$Sn$(d,p)$ & $1h_{11/2}$  & $33.6\%$ & $7.7\%$ & $26.4$           & $35^\circ$ \\  
$^{132}$Sn$(d,p)$ & $2f_{7/2}$  &$3.2\%$    & $2.5\%$ & $4.2\%$            & $16^\circ$ \\ 
$^{208}$Pb$(d,p)$ & $2g_{9/2}$  &$35.0\%$   & $12.6\%$ & $20.5\%$         & $32^\circ$ \\ 
\hline
\end{tabular}
\caption{Percent difference of the $(d,p)$ transfer cross sections at the first peak when using nonlocal potentials in entrance and exit channels (1st column), nonlocal potentials in entrance channel only (2nd column), and nonlocal potentials in exit channel only (3rd column), relative to the local calculation with the LPE potentials, for a number of reactions occurring at 
$20$ MeV.}
\label{Tab:Percent_Difference_20}
\end{table}

At higher energies, the reaction becomes more sensitive to the interior region, and consequently the reduction of the cross section due to the proton scattering state becomes more significant. At these higher energies, there is a competition between the two processes since there is contribution from both the periphery and the internal region. It was found that nonlocality in the proton scattering state had a larger effect for the heavier nuclei due to a larger surface region being probed. Comparing the dot-dashed green line with the dotted blue line of Fig. \ref{fig:TransferCS_HighE}, we can see that there is still enhancement of the cross section for $^{48}$Ca, but a reduction in $^{132}$Sn and $^{208}$Pb. The net effect of nonlocality in the exit channel depends on a complex interplay of the properties of the bound state (i.e. number of nodes, binding energy, and orbital angular momentum), and the magnitude of the real and imaginary part of the scattering wave function near the nuclear surface, where the majority of the transfer process occurs.

Now we consider the effects of nonlocality in the deuteron scattering state. Like the proton scattering state, nonlocality in the deuteron channel reduces the amplitude of the scattering wave function within the range of the nuclear interaction. However, unlike the proton scattering state, nonlocality in the deuteron channel also has the effect of shifting the scattering wave function outward. Both of these effects are evident in Fig. \ref{fig:dWF_Comparison}. Comparing the red dashed line to the blue dotted line in Fig. \ref{fig:TransferCS_LowE} and Fig. \ref{fig:TransferCS_HighE}, we see that for $^{48}$Ca,  nonlocality in the deuteron scattering state  has a similar effect as nonlocality in the proton scattering state, namely it reduces the cross section. As the size of the target increases, the  outward shift of the wave function becomes more important and increases the transfer cross section. This effect becomes largest for the heaviest targets. 

As seen in the insets of Fig. \ref{fig:TransferCS_LowE} where the predicted theoretical cross sections are normalized to the data at the peak of the distribution, the low energy data cannot distinguish between the various models since the shapes of the distributions are rather similar.  This is not the case for the higher energy cases we studied. For example for $^{48}$Ca(d,p) at 50 MeV, nonlocality significantly improves the description of the data. In all cases, if one is to extract a spectroscopic factor from these data, the results including nonlocal interactions explicitly will differ significantly from those where only local interactions are assumed.

The overall picture is compiled in Tables \ref{Tab:Percent_Difference_10}, \ref{Tab:Percent_Difference_20} and \ref{Tab:Percent_Difference_50}.
We show the percentage difference between cross sections including nonlocal and local interactions, at the peak of the angular distribution, relative to the local cross section. In the first column we include nonlocality in all nucleon target interactions, and in the second (third) column we include nonlocality in the entrance (exit) channel only.
At low energies (Tables \ref{Tab:Percent_Difference_10}, \ref{Tab:Percent_Difference_20}), the effect of nonlocality in the deuteron scattering state has a moderate impact on the transfer cross section. For these cases, the enhancement of the cross section due to the nonlocality in the exit channel is the dominant effect. This is also the case for (d,p) on light nuclei at 50 MeV (see Table \ref{Tab:Percent_Difference_50}). It is for the (d,p) reactions on heavy targets ($^{132}$Sn and $^{208}$Pb) at 50 MeV that nonlocality in the deuteron channel produces a substantial increase in the cross section.  

\begin{table}[h]
\centering
\begin{tabular}{|c|r|r|r|r|r|}
\hline
                 &      &                   & Nonlocal           & Nonlocal    &         \\       
                 &Final & Nonlocal          & Entrance           & Exit        &  Angle       \\
$E_{lab}$         &Bound & Relative          & Relative           & Relative    &  of  \\
$50$ MeV         &State & to Local          & to Local           & to Local    &  Peak \\
\hline
$^{16}$O$(d,p)$   & $1d_{5/2}$  & $22.3\%$  & $3.1\%$ & $15.8\%$  & $10^\circ$  \\ 
$^{16}$O$(d,p)$   & $2s_{1/2}$  & $20.7\%$  & $0.4\%$ & $21.2\%$  & $0^\circ$  \\ 
$^{40}$Ca$(d,p)$  & $1f_{7/2}$  & $4.8\%$  & $4.4\%$ & $0.2\%$             & $0^\circ$  \\ 
$^{48}$Ca$(d,p)$  & $2p_{3/2}$  & $41.9\%$  & $-8.1\%$ & $39.9\%$          & $0^\circ$  \\ 
$^{126}$Sn$(d,p)$ & $1h_{11/2}$ & $6.9\%$  & $6.7\%$ & $-2.5$            & $13^\circ$  \\  
$^{132}$Sn$(d,p)$ & $2f_{7/2}$  & $-10.9\%$  & $20.4\%$ & $-26.2\%$       & $0^\circ$  \\ 
$^{208}$Pb$(d,p)$ & $2g_{9/2}$  & $64.8\%$  & $86.5\%$ & $-1.7\%$         & $0^\circ$  \\ 
\hline
\end{tabular}
\caption{Percent difference of the $(d,p)$ transfer cross sections at the first peak when using nonlocal potentials in entrance and exit channels (1st column), nonlocal potentials in entrance channel only (2nd column), and nonlocal potentials in exit channel only (3rd column), relative to the local calculation with the LPE potentials, for a number of reactions occurring at 
$50$ MeV.}
\label{Tab:Percent_Difference_50}
\end{table}


\section{Discussion}
\label{Comparisons}

\subsection{Comparing DWBA and ADWA}
\label{DWBA_vs_ADWA}

The first order distorted wave Born approximation (DWBA) is still a common technique used in the analysis of transfer cross sections. It is based on a series expansion which, when truncated to first order as its usual implementation, only accounts for deuteron breakup  through its effects on deuteron elastic scattering. Unlike the adiabatic approach, built on a three-body model using only nucleon optical potentials,  DWBA uses a deuteron optical potential traditionally fit to deuteron elastic scattering. Although an adiabatic approximation is made in ADWA, no perturbative expansion is included and indeed deuteron breakup is included to all orders in the problem. As it turns out, ADWA provides a reliable solution to the three-body problem when compared to exact Faddeev calculations \cite{Nunes_prc2011}. As we will show in this section,  the differences in the DWBA and ADWA formalism can lead to very different predictions for the $(d,p)$ cross sections.

\begin{figure}[h]
\begin{center}
\includegraphics[width=0.5\textwidth]{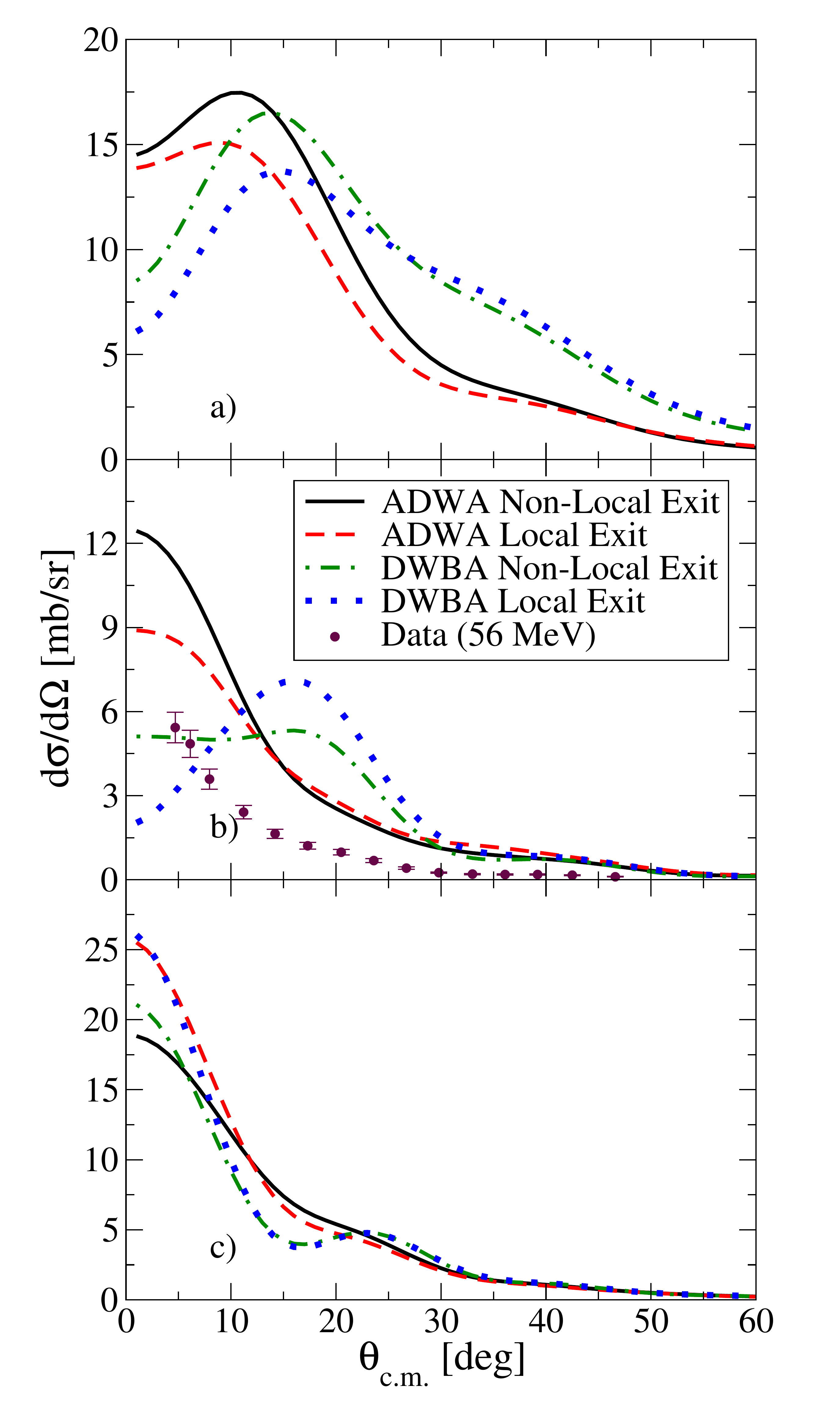}
\end{center}
\caption{Comparison of $(d,p)$ transfer cross sections when using ADWA as compared to DWBA. (a) $^{16}$O$(d,p)^{17}$O (b) $^{48}$Ca$(d,p)^{49}$Ca with data \cite{Uozumi_npa1994} at $56$ MeV. (c) $^{132}$Sn$(d,p)^{132}$Sn. All distribution at $E_d=50$ MeV.}
\label{fig:ADWA_vs_DWBA}
\end{figure}

In Fig. \ref{fig:ADWA_vs_DWBA} the angular distributions for three different (d,p) reactions obtained within DWBA are compared to those obtained with ADWA. We performed calculations including only local interactions, but also calculations where nonlocality is included in the exit channel. We found no meaningful way of comparing the effect of nonlocality in the entrance channel since DWBA and ADWA treat the deuteron channel very differently and a global nonlocal deuteron optical potential is not available. For a useful comparison,  the same nonlocal and local potentials are used in the exit channel. For the ADWA calculations, the LPE potentials discussed earlier were used to construct the adiabatic potential, while in the DWBA calculations the deuteron optical potential of Daehnick was used \cite{Daehnick_prc1980}. 

First we focus on the local calculations, and compare in Fig. \ref{fig:ADWA_vs_DWBA}, DWBA (dotted blue line) with ADWA (dashed red line). The shapes are significantly different, in addition to different magnitudes at the peak of the distribution. Including nonlocality in the exit channel does not change this picture. We compare DWBA with nonlocality in the exit channel (green dot-dashed line) with the corresponding ADWA (solid black line).  Qualitatively, the effect of nonlocality manifests in a similar way in DWBA and ADWA, producing an increase of the cross section at lower energies, an effect reduced or reverted at higher energies. Also shown in  Fig. \ref{fig:ADWA_vs_DWBA}(b) are the data. It is clear that for this case DWBA cannot describe the angular distribution from experiment. This is one example that demonstrates the need to include deuteron breakup in the reaction explicitly.

\subsection{Energy Shift}
\label{Energy_Shift}

Given that coordinate based theories are simpler to solve for local interactions,  Timofeyuk and Johnson \cite{Timofeyuk_prl2013, Timofeyuk_prc2013} developed a method to include the effects of nonlocality in the deuteron scattering state without having to solve the nonlocal equations. This method is based on the zero-range ADWA, and assumes the Perey-Buck form for the nonlocal potential. The main conclusion of the work is that, by shifting the energy by about $\sim 40$ MeV in the evaluation of the nucleon optical potentials from the standard $E_d/2$ value, one can capture the effects of nonlocality. We would like to test this method by comparing it to results when nonlocality is taken explicitly into account.

To reduce the ambiguities in the test, we fix the proton exit channel. 
For the nonlocal calculations,  in the exit channel, the LPE potentials found from fits to Perey-Buck proton elastic scattering distributions are used, along with the local binding potential used to reproduce the experimental neutron bound state. In the entrance channel, we use the nonlocal Perey-Buck potential, and the corresponding LPE potentials. For the method of \cite{Timofeyuk_prl2013,Timofeyuk_prc2013}, we use the CH89  potential \cite{ch89} evaluated at the standard $E_d/2$, with the additional energy shift quantified in \cite{Timofeyuk_prc2013}, and compare it to a local calculation using CH89 without any energy shift.

\begin{figure}[h]
\begin{center}
\includegraphics[width=0.5\textwidth]{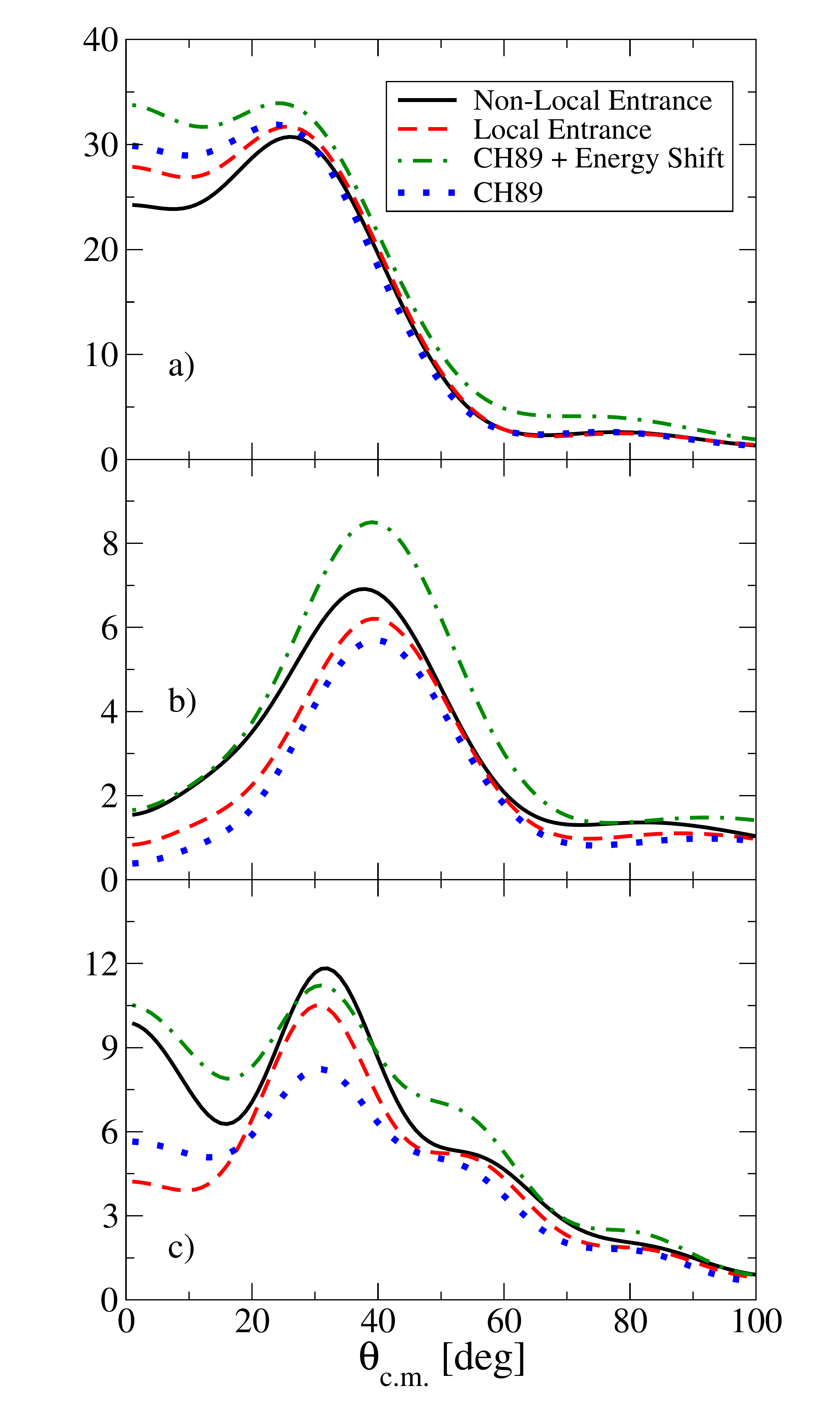}
\end{center}
\caption{Comparison of $(d,p)$ transfer cross sections with the explicit inclusion of nonlocality in the deuteron scattering state, and when using the proposed energy shift to take nonlocality in the deuteron scattering state into account. (a) $^{16}$O$(d,p)^{17}$O at $E_d=10$ MeV (b) $^{40}$Ca$(d,p)^{41}$Ca at $E_d=10$ MeV (c) $^{208}$Pb$(d,p)^{209}$Pb at $E_d=20$ MeV.}
\label{fig:Energy_Shift}
\end{figure}

The results of this study are shown in Fig. \ref{fig:Energy_Shift}. Angular distributions for (d,p) reactions on (a) $^{16}$O at $E_d=10$ MeV, (b) $^{40}$Ca at $E_d=10$ MeV and (c) $^{208}$Pb at $E_d=20$ MeV are shown: the black solid line corresponds to the explicit inclusion of nonlocality in the deuteron scattering state, the red dashed line are the results of the local calculation when the LPE potentials are used for the deuteron scattering state, the green dot dashed line is for the results using the method of \cite{Timofeyuk_prl2013,Timofeyuk_prc2013}, and the blue dotted line is when they were evaluated at the standard $E_d/2$ value. 

Our results demonstrate that the energy shift method always results in an increase in the cross section. As we saw in Sec/ \ref{Transfer}, explicit inclusion of nonlocality in the deuteron scattering state can produce an increase in the cross section, but this is not always the case (e.g. see Fig. \ref{fig:Energy_Shift}(a). The energy shift often times moves the transfer distribution towards larger angles, an effect also present when nonlocal interactions are accounted for exactly. However,  sometimes the energy shift method overshoots this effect, as in Fig. \ref{fig:Energy_Shift}b. An example when the energy shift method did a good job at reproducing the nonlocal effects is seen in Fig \ref{fig:Energy_Shift}c. In general, we found that, in most cases, the energy shift captured the qualitative effects of nonlocality, but did not provide an accurate account of the process. 

\subsection{Sensitivity to Binding Energy and Angular Momentum}
\label{Sensitivity}

We investigated the sensitivity of our results on the binding energy and the angular momentum of the final bound state. We find that the higher the binding energy the more the ANC changes when nonlocality is included, relative to the ANC of the bound state resulting from local potentials. For low energy reactions, which are dominated by the ANC, it enhances the effects we see in this study. For higher energy reactions, there is an interplay between the effects of the ANC of the bound state and the effects of nonlocality on the proton and deuteron scattering state. In general, the magnitude of the percent difference of the cross section at the first peak was more significant for higher binding, but no clear trend was observed. For low energy deuterons, we observe that nonlocality in the deuteron scattering state was more significant for higher binding. We also investigated the dependence on the angular momentum of the final bound state, but no obvious trend could be found for the limited cases we studied. 


\section{Conclusions}
\label{conclusions}

In this work we studied the effects of nonlocality on $(d,p)$ transfer reactions. An extension of the ADWA theory was developed to include nonlocality in the deuteron scattering state  using the Perey-Buck nonlocal nucleon optical potential \cite{Perey_np1962}. In the exit channel the Perey-Buck potential was used to describe the proton scattering state, and its real part was adjusted for the neutron bound state. For the scattering, NA interaction, a local phase equivalent potential was obtained by fitting the elastic scattering generated from the corresponding nonlocal potential. Both the local and nonlocal bound states reproduced the same experimental binding energies. 

For the $(d,p)$ reactions studied, we found that the inclusion of nonlocality in both the entrance and exit channels increased the transfer cross section by $\sim 40\%$. In most cases, nonlocality in the deuteron scattering state  caused a moderate increase in the transfer cross section. However, for heavy targets at high energies, this increase was large.  Nonlocality in the exit channel caused, almost exclusively,  an increase in the transfer cross section, except for heavy targets at high energies for which the cross sections were reduced. We also compared our ADWA result with those from DWBA and found the effects of non locality in the final state to be consistent in both formulations, even if quantitatively different. We also compared our ADWA results with the energy shift method introduced by Timofeyuk and Johnson \cite{Timofeyuk_prl2013,Timofeyuk_prc2013} and found that method to be qualitatively correct.

The conclusion of the present study confirm those of \cite{Titus_prc2014,Ross_prc2015}.  There are important differences in the transfer cross sections when including nonlocality explicitly as compared to when using local phase equivalent potentials. This highlights the necessity of explicitly including nonlocality to describe transfer reactions. 


\begin{center}
\textbf{ACKNOWLEDGMENTS}
\end{center}

We are grateful to Ian Thompson and Ron Johnson for 
many useful suggestions and invaluable advice.
This work was supported by the National Science
Foundation under Grants No. PHY-1068571 and PHY-1403906 and the
Department of Energy under Contracts No. DE-FG52-
08NA28552.


\appendix


\section{Nonlocal Adiabatic Potential}
\label{Derivation}

\noindent Here we will derive the partial wave decomposition of the nonlocal adiabatic potential. For illustration purposes, consider just the neutron case first. The neutron nonlocal operator acting on the three-body wave function is given by

\begin{eqnarray}
\hat{U}_{nA}\Psi(\textbf{r},\textbf{R})&=&\int U_{nA}(\textbf{R}_n,\textbf{R}'_n)\Psi(\textbf{R}'_n,\textbf{R}'_p) \nonumber \\
&\phantom{=}& \times \ \delta(\textbf{R}'_p-\textbf{R}_p)d\textbf{R}'_nd\textbf{R}'_p \nonumber \\
&=&8\int U_{nA}\left(\textbf{R}-\frac{\textbf{r}}{2},2\textbf{R}'-\textbf{R}-\frac{\textbf{r}}{2}\right) \nonumber \\
&\phantom{=}& \times \ \Psi(\textbf{r}-2(\textbf{R}'-\textbf{R}),\textbf{R}')d\textbf{R}'.
\end{eqnarray}

\noindent The Jacobian for the coordinate transformation is unity, and we integrated over $d\textbf{r}'$ to eliminate the delta function. We used the vector definitions $\textbf{R}_{p,n}=\textbf{R}\pm\frac{\textbf{r}}{2}$, where $\textbf{R}_p$ uses the ``$+$'' sign and $\textbf{R}_n$ uses the ``$-$'' sign. A similar expression is found for the proton nonlocal operator. We expand our wave function in a basis of Weinberg states, as in Eq.(\ref{eq-psi}). Since we are using only the first Weinberg state, $\Psi(\textbf{r},\textbf{R})\approx \Phi(\textbf{r})X(\textbf{R})$ and the nucleon nonlocal operator is

\begin{eqnarray}
&\phantom{=}&\hat{U}_{NA}\Phi(\textbf{r})X(\textbf{R})=8\int U_{NA}\left(\textbf{R}\pm\frac{\textbf{r}}{2} ,2\textbf{R}'-\textbf{R}\pm\frac{\textbf{r}}{2}\right) \nonumber \\
&\phantom{=}& \quad \quad \times \ \Phi(\textbf{r}\pm2(\textbf{R}'-\textbf{R}))X(\textbf{R}')d\textbf{R}'.
\end{eqnarray}

\noindent Adding and subtracting by $\textbf{R}$ in the second argument of $U_{NA}$ and making the definition $\textbf{s}=\textbf{R}'-\textbf{R}$, we can rewrite the nucleon nonlocal operator as

\begin{eqnarray}
\hat{U}_{NA}\Phi(\textbf{r})X(\textbf{R})&=&8\int U_{NA}\left(\textbf{R}_{p,n},\textbf{R}_{p,n}+2\textbf{s}\right) \nonumber \\
&\phantom{=}& \times \ \Phi(\textbf{r}\pm 2\textbf{s})X(\textbf{R}+\textbf{s})d\textbf{s}.
\end{eqnarray}

\noindent The general expansion of a partial wave with total angular momentum and projection, $J_T M_T$, is given by:

\begin{eqnarray}\label{WFexpansion}
&\phantom{=}& \Psi(\textbf{r},\textbf{R})\approx \Phi(\textbf{r})X(\textbf{R})=\sum_{\ell L J_p}\phi_{\ell}(r)\frac{\chi_{L J_p}^{J_T M_T}(R)}{R} \\
&\times& \left\{ \left\{ \left\{ \left\{ \Xi_{1/2}(\xi_n)\otimes \Xi_{1/2}(\xi_p)\right\}_1    \otimes \tilde{Y}_{\ell}(\hat{r})\right\}_{1} \otimes \tilde{Y}_{L}(\hat{R})\right\}_{J_p} \right. \nonumber \\
&\times& \ \left. \otimes \Xi_{I_t}(\xi_t) \right\}_{J_T M_T}. \nonumber
\end{eqnarray}

\noindent In this equation, $\Xi_{1/2}(\xi_{n})$ and $\Xi_{1/2}(\xi_{p})$ are the spin functions for the neutron and proton respectively, and $\Xi_{I_t}(\xi_t)$ is the spin function for the target, with $I_t$ being the spin of the target. The spin of the deuteron is coupled to the orbital angular momentum, $L$, to give the total angular momentum of the projectile, $J_p$. The total angular momentum of a given partial wave is given by $J_T$. $\tilde{Y}_{\ell}(\hat{r})$ and $\tilde{Y}_L(\hat{R})$ are spherical harmonics for the internal and relative orbital angular momentum of the deuteron, respectively. We use the phase convention where the spherical harmonics have a built in factor of $i^\ell$ so that $\tilde{Y}_{\ell}(\hat{r})=i^{\ell}Y_{\ell}(\hat{r})$, and similarly for $\tilde{Y}_L(\hat{R})$. 

We would like to find the partial wave decomposition of 

\begin{eqnarray}\label{NLeqn}
\left[\hat{T}_{\textbf{R}}-E_d \right]\Phi(\textbf{r})X(\textbf{R})=-\left(\hat{U}_{nA}+\hat{U}_{pA} \right)\Phi(\textbf{r})X(\textbf{R}). \nonumber \\
\end{eqnarray}

\noindent To do this, we multiply by

\begin{eqnarray}
&\phantom{=}&\sum_{\ell'}\phi_{\ell'}(r)V_{np}(r) \ \times \\
&\phantom{=}&\left\{ \left\{ \left\{\left\{ \Xi_{1/2}(\xi_n)\otimes \Xi_{1/2}(\xi_p)\right\}_1 \otimes \tilde{Y}_{\ell'}(\hat{r})\right\}_{1} \otimes \tilde{Y}_{L'}(\hat{R})\right\}_{J'_p} \right. \nonumber \\
&\phantom{=}& \times \ \left. \otimes \Xi_{I_t}(\xi_t) \right\}_{J_T M_T}^* \nonumber
\end{eqnarray}

\noindent and integrate over $d\textbf{r}$, $d\Omega_R$, $d\xi_{n}$, $d\xi_{p}$ and $d\xi_t$. The $l.h.s.$ of the equation becomes

\begin{eqnarray}\label{lhs}
\frac{1}{R}\left[\frac{\hbar^2}{2\mu}\left(\frac{\partial^2}{\partial R^2}-\frac{L'(L'+1)}{R^2} \right)+E_d \right]\chi_{L' J'_p}^{J_T M_T}(R).
\end{eqnarray}

\noindent As we only considered $\ell=0$ deuterons in our calculations, let us make this assumption right at the beginning of our partial wave decomposition of the $r.h.s.$  Therefore, the two $\tilde{Y}_{\ell}(\hat{r})$ terms give $1/4\pi$, and the partial wave decomposition of the $r.h.s.$ of Eq.(\ref{NLeqn}) is:

\begin{eqnarray}
&\phantom{=}& -\frac{8}{4\pi}\sum_{L J_p} \int \phi_{0}(r)V_{np}(r) U_{NA}\left(\textbf{R}_{p,n},\textbf{R}_{p,n}+2\textbf{s}  \right) \nonumber \\
&\times& \left\{ \left\{ \left\{ \Xi_{1/2}(\xi_n)\otimes \Xi_{1/2}(\xi_p)\right\}_1 \otimes \tilde{Y}_{L'}(\hat{R})\right\}_{J'_p} \right. \nonumber \\
&\times& \left. \otimes \Xi_{I_t}(\xi_t) \right\}_{J_T M_T}^* \nonumber \\
&\times& \left\{ \left\{ \left\{ \Xi_{1/2}(\xi_n)\otimes \Xi_{1/2}(\xi_p)\right\}_1 \otimes \tilde{Y}_{L}(\widehat{R+s})\right\}_{J_p} \right. \nonumber \\
&\times& \left. \otimes \Xi_{I_t}(\xi_t) \right\}_{J_T M_T} \\
&\times& \phi_{0 }(|\textbf{r}\pm 2\textbf{s}|)\frac{\chi_{L J_p}^{J_T M_T}\left(\left|\textbf{R}+\textbf{s} \right|\right)}{\left|\textbf{R}+\textbf{s} \right|} d\textbf{s} d\textbf{r} d\Omega_R d\xi_t d\xi_{n}d\xi_{p} \nonumber
\end{eqnarray}

Our goal is to couple the integrand up to zero angular momentum. Therefore, it will be spherically symmetric so we can use symmetry to reduce the dimensionality of the integral. Coupling the spherical harmonics together and the spin functions together, then doing the integral over $d\xi_n$, $d\xi_p$ and $d\xi_t$, we arrive at

\begin{eqnarray}
&\phantom{=}&-\frac{8}{4\pi\hat{L}'}\int \phi_0(r)V_{np}(r) U_{NA}\left(\textbf{R}_{p,n},\textbf{R}_{p,n}+2\textbf{s}  \right) \nonumber \\
&\phantom{=}& \times \ \left\{\tilde{Y}_{L'}(\hat{R}) \otimes \tilde{Y}_{L'}(\widehat{R+s})\right\}_{00}  \phi_{0 }(|\textbf{r}\pm 2\textbf{s}|) \nonumber \\
&\phantom{=}& \times \ \frac{\chi_{L' J'_p}^{J_T M_T}\left(\left|\textbf{R}+\textbf{s} \right|\right)}{\left|\textbf{R}+\textbf{s} \right|} d\textbf{s} d\textbf{r} d\Omega_R,  \nonumber \\
\end{eqnarray} 

\noindent where $\hat{L}'=\sqrt{2L'+1}$. We can bring the term $\frac{1}{R}$ from Eq.(\ref{lhs}) over to the $r.h.s.$ Since the integrand is coupled to zero angular momentum, it is spherically symmetric, which means that it is invariant under rotations of the three vectors $\textbf{R}$, $\textbf{r}$, and $\textbf{s}$ together. Thus, we can evaluate it in any configuration we want. By placing the $\textbf{R}$ vector in the $\hat{z}$-direction, $M=0$, and $\tilde{Y}_{L0}(\hat{z})=\frac{i^L\hat{L}}{\sqrt{4\pi}}$. We will place the vector $\textbf{r}$ in the $xz$-plane so that the $\phi_r$-dependence is removed. Integration over $d\Omega_R$ yields a factor of $4\pi$ for all other choices of which direction to orient the vector $\textbf{R}$. Since we are fixing $\textbf{r}$ to be in the $xz$-plane, we get a factor of $2\pi$ to take care of rotations around the z-axis. Thus, we can evaluate our integral in this specific configuration, then multiply the integral by $8\pi^2$ to take care of all other configurations. There is not enough symmetry to fix the vector $\textbf{s}$. After explicitly adding in the $i^L$ phase from the remaining spherical harmonic, and redefining all of our angular momentum indices in terms of unprimed quantities, we get

\begin{eqnarray}
&\phantom{=}&\left[\frac{\hbar^2}{2\mu}\left(\frac{\partial^2}{\partial R^2}-\frac{L(L+1)}{R^2} \right)+E_d \right]\chi_{L J_p}^{J_T M_T}(R) \\
&=&-\frac{8R \sqrt{\pi}}{\hat{L}}\int \phi_0(r)V_{np}(r) U_{NA}\left(\textbf{R}_{p,n},\textbf{R}_{p,n}+2\textbf{s}  \right)Y_{L0}(\widehat{R+s})   \nonumber \\
&\phantom{=}& \times \ \phi_{0}(|\textbf{r}\pm 2\textbf{s}|)\frac{\chi_{L J_p}^{J_T M_T}\left(\left|\textbf{R}+\textbf{s} \right|\right)}{\left|\textbf{R}+\textbf{s} \right|}r^2 \sin\theta_{r} dr d\theta_{r}d\textbf{s}, \nonumber
\end{eqnarray}

\noindent where $\textbf{R}_{p,n}$, $\textbf{R}$, and $\textbf{s}$ are to be evaluated in the particular configuration described above. $U_{NA}$ is the nucleon optical potential for either the proton or the neutron. Making the replacement $U_{NA}\rightarrow U_{nA}+U_{pA}$ gives us the full nonlocal adiabatic potential, and the resulting partial wave equation for the deuteron scattering state when using nonlocal potentials within the ADWA.


\bibliography{nonlocal_transfer}

\end{document}